\documentclass[usegraphicx,usenatbib]{mn2e}

\date{\today}

\begin{document}

\twocolumn

\title[Adiabatic vs Isocurvature Non--Gaussianity] {Adiabatic versus
Isocurvature Non--Gaussianity}

\author[Hikage et al.]
{Chiaki Hikage$^1$, Dipak Munshi$^2$, Alan Heavens$^2$ and Peter Coles$^3$ \\
$^1$ Department of Astrophysical Sciences, Princeton University, Peyton Hall, Princeton NJ 08544, USA \\
$^2$ Scottish Universities Physics Alliance (SUPA),~ Institute for Astronomy, University of Edinburgh, Blackford Hill,  Edinburgh EH9 3HJ \\
$^3$ School of Physics and Astronomy, Cardiff University, Cardiff, CF24 3AA \\
}
\maketitle 

\begin{abstract}
We study the extent to which one can distinguish primordial
non--Gaussianity (NG) arising from adiabatic and isocurvature
perturbations.  We make a joint analysis of different NG models based
on various inflationary scenarios: {\em local-type} and {\em
equilateral-type} NG from adiabatic perturbations and {\em local-type}
and {\em quadratic-type} NG from isocurvature perturbations together
with a foreground contamination by point sources. We separate the
Fisher information of the bispectrum of CMB temperature and
polarization maps by $l$ for the {\em skew spectrum} estimator
introduced by \citet{MunshiHeavens} to study the scale dependence of
the signal-to-noise ratio of different NG components and their
correlations. We find that the adiabatic and the isocurvature modes
are strongly correlated, though the phase difference of acoustic
oscillations helps to distinguish them. The correlation between local-
and equilateral-type is weak, but the two isocurvature modes are too
strongly correlated to be discriminated. Point source contamination,
to the extent to which it can be regarded as white noise, can be
almost completely separated from the primordial components for
$l>100$.  Including correlations among the different components, we
find that the errors of the NG parameters increase by 20-30\% for the
WMAP 5-year observation, but $\simeq 5 \%$ for Planck observations.
\end{abstract}
\section{Introduction}

The statistical properties of fluctuations in the early Universe can
be used to probe the very earliest stages of its history, and provide
valuable information on the mechanisms which ultimately gave rise to
the existence of structure within it. This may include evidence for
the cosmic inflationary expansion. With the recent claim of a
detection of non--Gaussianity \citep{YaWa08} in the Wilkinson
Microwave Anisotropy Probe (WMAP) sky maps, interest in
primordial non--Gaussianity has obtained a tremendous boost.

Non-Gaussianity from the simplest inflationary models based on a
single slowly-rolling scalar field is typically very small
\citep{Salopek90,Falk93,Gangui94,Acq03,Mal03,Bartolo06}. Variants of
the simple inflationary models can lead to much higher levels of
non--Gaussianity, such as multiple fields \citep{LM1997,Lyth03};
modulated reheating scenarios \citep{DGZ2004}; warm inflation
\citep{GuBeHea02,Moss}; ekpyrotic model \citep{Koya07}.

Different forms are proposed to describe primordial non--Gaussianity.
Much interest has focused on {\em local-type} $f_{\rm NL}$ by which
the non--Gaussianity of Bardeen's curvature perturbations is locally
characterized \citep{Gangui94,Verde00,WangKam00,KomSpe01,BZ04}:
\begin{equation}
\label{eq:adiloc}
\Phi(x) = \phi(x) + f_{\rm NL}(\phi^2(x) - \langle
\phi^2(x) \rangle ),
\end{equation}
where $\phi$ is the linear Gaussian part of $\Phi$. This form is
motivated by the single-field inflation scenarios and then many models
predict non--Gaussianity in terms of $f_{\rm NL}$ \citep{BKMR04}.
Optimized estimators of the bispectrum, which is the leading
correlation term in the local form, are introduced by \cite{Heav98}
and have been successively developed to the point where an estimator
for $f_{\rm NL}$ saturates the Cram\'{e}r-Rao bound for partial sky
coverage and inhomogeneous noise
\citep{KSW,Crem06,Crem07b,MedeirosContaldi06,Cabella06,Liguori07,Komatsu09,SmSeZa09}.

The local-type $f_{\rm NL}$ is sensitive to the bispectrum with
squeezed-configuration triangles ($k_1\ll k_2 \sim k_3$).  Several models
including the inflation scenario with non-canonical kinetic terms
\citep{SL05,Chen07b}, Dirac-Born-Infeld models \citep{Ali04}, and Ghost
inflation \citep{Ark04} predict large NG signals in
equilateral configuration triangles ($\ell_1\simeq \ell_2 \simeq
\ell_3$), which is well described with {\em equilateral-type} $f_{\rm
NL}$ \citep{BCZ04}.

Non-Gaussianity arising from primordial isocurvature (entropy)
perturbations has been discussed in the context of NG field potentials
\citep{LM1997,Peebles1999,BL2006,ST2008}, the curvaton scenario
\citep{Lyth03,BMR2004,Beltran2008,MT2009}, modulated reheating
\citep{BC2006}, baryon asymmetry \citep{KNT2009}, and the axion
\citep{KNSST2008}.  \citet{HKMTY2009} first put observational limits on
the isocurvature non-Gaussianity using WMAP 5-year data.

In this paper, we make a joint analysis of the different NG models to
estimate the extent to which one can decode each NG information from
CMB temperature (T) and E polarization (E) maps obtained by WMAP and
Planck.  We separate Fisher information of the CMB bispectrum by
different ranges of $l$ to study at which angular scale each NG
parameter has large S/N and correlations among different NG components
weaken.  This idea is based on a new estimator called {\em skew
spectrum}, which \citet{MunshiHeavens} has introduced to measure a
scale dependence of NG parameters, while the commonly-used single
skewness parameter \citep{KSW} gives a single value averaged over all
scales.  The advantage of the new estimator is that it retains
information on the source of the non-Gaussianity, which the
commonly-used one does not.

For our analysis, we adopt a set of cosmological parameters at the
maximum likelihood values for a power-law $\Lambda$CDM model from the
WMAP 5-year data only fit \citep{Dunkley2009}: $\Omega_{\rm
b}=0.0432$; $\Omega_{\rm cdm}=0.206$; $\Omega_\Lambda=0.7508$;
$H_0=72.4~{\rm km~s^{-1}~Mpc^{-1}}$; $\tau=0.089$; $n_\phi=0.961$.
The amplitude of the primordial power spectrum is set to be
$2.41\times 10^{-9}$ at $k=0.002{\rm Mpc}^{-1}$. The
spectra of isocurvature perturbations are assumed to be
scale-invariant. The radiation transfer functions for adiabatic and
isocurvature perturbations are computed using the publicly-available
CMBFAST code \citep{SZ1996}.

This paper is organized as follows: different NG models from
primordial adiabatic and isocurvature perturbations are introduced in
\S \ref{sec:model}; \S \ref{sec:fisher} presents a Fisher matrix
analysis of these parameters in which we estimate the corresponding
error expected from WMAP and Planck observations; \S \ref{sec:summary}
devotes to a summary.

\section{Models of Primordial Non-Gaussianity}
\label{sec:model}

We consider various forms to describe primordial non--Gaussianity
from adiabatic and isocurvature perturbations, and then
provide explicit expressions for the bispectra.

\subsection{Local-Type Adiabatic component}
The bispectrum in the {\em local-type} NG form (eq.[\ref{eq:adiloc}])
is written as \citep[e.g.,][]{KomSpe01}
\begin{eqnarray}
\label{eq:badiloc}
B^{\rm Adi,Loc}(k_1,k_2,k_3)=2f_{\rm NL}^{\rm Adi,Loc}
[P_\phi(k_1)P_\phi(k_2)~~~~~~~~~~~~~ \nonumber \\
+P_\phi(k_2)P_\phi(k_3)+P_\phi(k_3)P_\phi(k_1)],
\end{eqnarray}
where we rewrite $f_{\rm NL}$ in the equation (\ref{eq:adiloc}) as
$f_{\rm NL}^{\rm Adi,Loc}$.  The CMB angular bispectra for $T$, $E$, and
their cross terms are given by
\begin{eqnarray}
b_{XYZ,l_1l_2l_3}^{\rm Adi,Loc}=2f_{\rm NL}^{\rm Adi,Loc}\int r^2dr[
\beta_{Xl_1}^{\rm Adi}(r)\beta_{Yl_2}^{\rm Adi}(r)\alpha_{Zl_3}^{\rm Adi}(r)
\nonumber \\
+\beta_{Xl_1}^{\rm Adi}(r)\alpha_{Yl_2}^{\rm Adi}(r)\beta_{Zl_3}^{\rm Adi}(r)
+\alpha_{Xl_1}^{\rm Adi}(r)\beta_{Yl_2}^{\rm Adi}(r)\beta_{Zl_3}^{\rm Adi}(r)]
\end{eqnarray}
where $X, Y$, and $Z$ denote $T$ or $E$, and $\alpha_{Xl}^{\rm Adi}$ and
$\beta_{Xl}^{\rm Adi}$ are defined with the adiabatic radiation transfer
function $g_{Xl}^{\rm Adi}$ as
\begin{eqnarray}
\alpha_{Xl}^{\rm Adi}(r)&\equiv &
\frac{2}{\pi}\int k^2dkg_{Xl}^{\rm Adi}(k)j_l(kr), \\
\beta_{Xl}^{\rm Adi}(r)&\equiv &
\frac{2}{\pi}\int k^2dkP_{\phi}(k)g_{Xl}^{\rm Adi}(k)j_l(kr).
\end{eqnarray}

\subsection{Equilateral-Type Adiabatic Component}
The bispectrum in the {\em equilateral-type} NG form is characterized 
by the NG parameter $f_{\rm NL}^{\rm Adi,Eq}$ \citep{BCZ04} as follows:
\begin{eqnarray}
B^{\rm Adi,Eq}(k_1,k_2,k_3)=6f_{\rm NL}^{\rm Adi,Eq}
[-P_\phi(k_1)P_\phi(k_2)-P_\phi(k_2)P_\phi(k_3)
\nonumber \\
-P_\phi(k_3)P_\phi(k_1)
-2\{P_\phi(k_1)P_\phi(k_2)P_\phi(k_3)\}^{2/3}
\nonumber \\
+\{[P_\phi(k_1)P_\phi(k_2)^2P_\phi(k_3)^3]^{1/3} + (5~{\rm perm.})\}].
\nonumber \\
\end{eqnarray}
The CMB angular bispectra in this form are given by
\begin{eqnarray}
b_{XYZ,l_1l_2l_3}^{\rm Adi,Eq} = 6f_{\rm NL}^{\rm Adi,Eq}\int r^2 dr 
[-\beta_{Xl_1}^{\rm Adi}(r)\beta_{Yl_2}^{\rm Adi}(r)\alpha_{Zl_3}^{\rm Adi}(r)
\nonumber \\
-\beta_{Xl_1}^{\rm Adi}(r)\alpha_{Yl_2}^{\rm Adi}(r)\beta_{Zl_3}^{\rm Adi}(r)
-\alpha_{Xl_1}^{\rm Adi}(r)\beta_{Yl_2}^{\rm Adi}(r)\beta_{Zl_3}^{\rm Adi}(r)
\nonumber \\
-2\delta_{Xl_1}^{\rm Adi}(r)\delta_{Yl_2}^{\rm Adi}(r)\delta_{Zl_3}^{\rm Adi}(r)
\nonumber \\
+\{\beta_{Xl_1}^{\rm Adi}(r)\gamma_{Yl_2}^{\rm Adi}(r)\delta_{Zl_3}^{\rm Adi}(r)+(5 {\rm perm.})\}],
\end{eqnarray}
where
\begin{eqnarray}
\gamma_{Xl}^{\rm Adi}(r)&\equiv &
\frac{2}{\pi}\int k^2dkP_{\phi}^{1/3}(k)g_{Xl}^{\rm Adi}(k)j_l(kr), \\
\delta_{Xl}^{\rm Adi}(r)&\equiv &
\frac{2}{\pi}\int k^2dkP_{\phi}^{2/3}(k)g_{Xl}^{\rm Adi}(k)j_l(kr).
\end{eqnarray}

\subsection{Isocurvature Components}
Here we consider an isocurvature perturbation ${\cal S}$ between
axion-type cold dark matter (CDM) and radiation, which is uncorrelated
with adiabatic perturbations, defined as
\begin{equation}
\label{eq:defiso}
{\cal S}\equiv\frac{\delta\rho_{\rm CDM}}{\rho_{\rm CDM}}
-\frac{3\delta\rho_\gamma}{4\rho_\gamma},
\end{equation}
where $\rho_{\rm CDM}$ is the CDM energy density and $\rho_\gamma$ is
the radiation energy density. The fractional isocurvature perturbation
$f_{\rm S}$ is defined as
\begin{equation}
\label{eq:fs}
f_{\rm S}\equiv\frac{P_{\cal S}(k_0)}{P_{\zeta}(k_0)+P_{\cal S}(k_0)}.
\end{equation}
where $P_\zeta$ and $P_{\cal S}$ represent the power spectra of
$\zeta$ and ${\cal S}$ and $k_0$ is set to be 0.002Mpc$^{-1}$. At
linear order, $\Phi$ (eq.[\ref{eq:adiloc}]) is related to $\zeta$ by
$\Phi=(3/5)\zeta$. The definition of $f_{\cal S}$ is same as the
commonly used parameter $\alpha$ \citep{BDP2006}.  The current
observational limit on $f_{\rm S}$ is 0.067 (95\% CL) for the
axion-type isocurvature perturbation \citep{Komatsu09}.

\subsubsection{Local-Type Isocurvature Component}
We consider two different forms for isocurvature non-Gaussianity.  One
is the same local form as the adiabatic one
(eq.[\ref{eq:badiloc}]):
\begin{eqnarray}
B^{\rm Iso,Loc}(k_1,k_2,k_3)=2f_{\rm NL}^{\rm Iso,Loc}
[P_\eta(k_1)P_\eta(k_2)~~~~~~~~~~~~~~ 
\nonumber \\
+P_\eta(k_2)P_\eta(k_3)+P_\eta(k_3)P_\eta(k_1)],
\end{eqnarray}
where $\eta$ denotes the Gaussian part of ${\cal S}$ with
the amplitude of $P_\eta(k)$ is normalized by $f_{\rm S}$
(eq.[\ref{eq:fs}]). The parameter $f_{\rm NL}^{\rm Iso,Loc}$
corresponds to $\alpha^2f_{\rm NL}^{\rm ISO}$ in \citet{HKMTY2009}.
We obtain the CMB bispectrum as
\begin{eqnarray}
\label{eq:bisiso1}
b_{XYZ,l_1l_2l_3}^{\rm Iso,Loc}=2f_{\rm NL}^{\rm Iso,Loc}\int r^2dr[
\beta_{Xl_1}^{\rm Iso}(r)\beta_{Yl_2}^{\rm Iso}(r)\alpha_{Zl_3}^{\rm Iso}(r)
\nonumber \\
+\beta_{Xl_1}^{\rm Iso}(r)\alpha_{Yl_2}^{\rm Iso}(r)\beta_{Zl_3}^{\rm Iso}(r)
+\alpha_{Xl_1}^{\rm Iso}(r)\beta_{Yl_2}^{\rm Iso}(r)\beta_{Zl_3}^{\rm Iso}(r)],
\end{eqnarray}
where $\alpha_{Xl}^{\rm Iso}$ and $\beta_{Xl}^{\rm Iso}$ are defined with
the isocurvature radiation transfer function $g_{Xl}^{\rm Iso}$ as
\begin{eqnarray}
\alpha_{Xl}^{\rm Iso}(r)&\equiv &\frac{2}{\pi}\int k^2dk
g_{Xl}^{\rm Iso}(k)j_l(kr), \\
\beta_{Xl}^{\rm Iso}(r)&\equiv &\frac{2}{\pi}\int k^2dkP_{\eta}(k)
g_{Xl}^{\rm Iso}(k)j_l(kr).
\end{eqnarray}

\subsubsection{Quadratic-Type Isocurvature Component}
When the linear Gaussian term is negligible compared with the
quadratic term, the isocurvature perturbation has a $\chi^2$ from
\citep[e.g.,][]{LM1997}:
\begin{equation}
\label{eq:isoquad}
{\cal S}=\sigma^2-\langle\sigma^2\rangle,
\end{equation}
where $\sigma$ obeys Gaussian statistics.  This form has been studied in the
context of axion \citep{KNSST2008} and curvaton scenarios
\citep{LVW2008}. The bispectra are calculated as \citep{Komatsu2002}
\begin{eqnarray}
\label{eq:bisquad}
B^{\rm Iso,Quad}(k_1,k_2,k_3)=
\frac{8}{3}\int_{L_{\rm box}^{-1}}\frac{d^3{\mathbf
p}}{(2\pi)^3}P_\sigma(p)~~~~~~~~~~~~~~~~~~ 
\nonumber \\
\times[P_\sigma(|{\mathbf k_1}+{\mathbf p}|)
P_\sigma(|{\mathbf k_2}-{\mathbf p}|)
+P_\sigma(|{\mathbf k_2}+{\mathbf p}|)
P_\sigma(|{\mathbf k_3}-{\mathbf p}|) \nonumber \\
+P_\sigma(|{\mathbf k_3}+{\mathbf p}|)
P_\sigma(|{\mathbf k_1}-{\mathbf p}|)],
\end{eqnarray}
where a finite box-size $L_{\rm box}$ gives an infrared cutoff.  To
avoid assumptions at scales far beyond the present horizon $H_0^{-1}$,
we set $L_{\rm box}=30$Gpc. The equation (\ref{eq:bisquad}) is
approximately given by \citet{HKMTY2009} as
\begin{eqnarray}
\label{eq:bisiso2}
b_{XYZ,l_1l_2l_3}^{\rm Iso,Quad}
=2 \int r^2dr
[\beta_{Xl_1}^{\rm Iso,Quad}(r)\beta_{Yl_2}^{\rm Iso}(r)\alpha_{Zl_3}^{\rm Iso}(r)~~~~~~~~~~~~~
\nonumber \\
+\beta_{Xl_1}^{\rm Iso}(r)\alpha_{Yl_2}^{\rm Iso}(r)\beta_{Zl_3}^{\rm Iso,Quad}(r)
+\alpha_{Xl_1}^{\rm Iso}(r)\beta_{Yl_2}^{\rm Iso,Quad}(r)\beta_{Zl_3}^{\rm Iso}(r)
], \nonumber \\
\end{eqnarray}
where
\begin{eqnarray}
\beta_{Xl}^{\rm Iso,Quad}(r) & 
\equiv & \frac{2}{\pi}\int_{L_{\rm box}^{-1}} k^2dk
P_{\cal S}(k)g_{Xl}^{\rm Iso}(k)j_l(kr), \\
\beta_{Xl}^{\rm Iso}(r) & \equiv & \frac{2}{\pi}\int_{L_{\rm box}^{-1}} k^2dk
P_\sigma(k)g_{Xl}^{\rm Iso}(k)j_l(kr).
\end{eqnarray}
Non-Gaussianity in this form is characterized by $f_{\rm S}$
(eq.[\ref{eq:fs}]).

\subsection{Point Source Component}
Unmasked point sources (e.g., radio galaxies) generates an additional
non--Gaussianity in observed CMB maps. Assuming them to be Poisson
distribution, $b_{XYZ,l_1l_2l_3}^{\rm PS}$ is a constant.

\section{Fisher Information Analysis for Skew Spectrum}
\label{sec:fisher}

We make Fisher information analysis of the different NG components
introduced in the previous section to estimate the error expected from
WMAP, Planck and noiseless ideal observations.

The Fisher matrix for the CMB bispectrum in the weakly non-Gaussian, 
all-sky limit is written as \citep{BZ04,YKW07}
\begin{eqnarray}
F^{ij}&=&\sum_{l} F_l^{ij}, \\
F_l^{ij}&=&\sum_{2\le l_1\le l_2\le l}I^2_{l_1l_2l}
\sum_{XYZ}\sum_{PQR}
\nonumber \\
&&\times b_{XYZ,l_1l_2l}^i
({\bf Cov}^{-1})_{l_1l_2l}^{XYZ|PQR}
b_{PQR,l_1l_2l}^j,
\label{eq:fisher}
\end{eqnarray}
where $i$ and $j$ denote each NG component and the factor
$I_{l_1l_2l_3}$ is defined as
\begin{equation}
I_{l_1l_2l_3}\equiv
\sqrt{\frac{(2l_1+1)(2l_2+1)(2l_3+1)}{4\pi}}
\left(
\begin{array}{ccc}l_1&l_2&l_3\\0&0&0\end{array}
\right).
\label{eq:I_def}
\end{equation}
The sums over $XYZ$ and $PQR$ are just $TTT$ when using CMB temperature
maps only (T only), but are eight combinations ($TTT, TTE, TET, ETT,
TEE, ETE, EET, EEE$) when both CMB temperature and E polarization
maps are used (T\&E).  The Fisher matrix at each $l$, $F_l^{ij}$, is
associated with the skew spectrum estimator for the $i$-th NG
component, $S_l^i$, defined as \citep{MunshiHeavens}
\begin{eqnarray}
S_l^i&=&\frac{1}{2l+1}\sum_{2\le l_1\le l_2\le l}
I^2_{l_1l_2l}\sum_{XYZ}\sum_{PQR} 
\nonumber \\
&&\times b_{XYZ,l_1l_2l}^i
({\bf Cov}^{-1})_{l_1l_2l}^{XYZ|PQR}
b_{PQR,l_1l_2l}^{\rm obs},
\end{eqnarray}
where $b^{\rm obs}_{l_1l_2l}$ denotes the observed bispectrum.  The
relation to the single skewness estimator $S_{\rm prim}^i$ \citep{KSW}
is
\begin{equation}
S_{\rm prim}^i=\sum_l(2l+1)S_l^i.
\end{equation}
When non--Gaussianity is small, the covariance matrix is approximately
given by
\begin{equation}
{\bf Cov}_{l_1l_2l_3}^{XYZ|PQR}\simeq \Delta_{l_1l_2l_3}
C_{l_1}^{XP}C_{l_2}^{YQ}C_{l_3}^{ZR},
\end{equation}
where $\Delta_{l_1l_2l_3}$ is 6 ($l_1=l_2=l_3$), 2 ($l_1=l_2$,
$l_2=l_3$, or $l_1=l_3$), and 1 ($l_1\ne l_2\ne l_3$) and $C_l^{XY}$
represents the CMB power spectrum from purely adiabatic perturbations
including observational noise $N_l^{XY}$:
\begin{equation}
C_l^{XY}=\frac{2}{\pi}\int k^2dk P_\phi(k)g_{Xl}^{\rm
Adi}(k)g_{Yl}^{\rm Adi}(k) + N_l^{XY}.
\end{equation}
We consider three different noise/beam functions: an ideal case
without noise/beam (``Ideal''); WMAP 5-year V+W band coadded map
(``WMAP5''); Planck's expectations after two full sky surveys for 14
months (``Planck'') using all of nine frequency channels. Noise is
assumed to be homogeneous white noise and $N_l^{XY}=0$ when $X\neq
Y$. Noise/beam is coadded at each $l$ with the inverse weight of the
noise variance in each frequency band or differential assembly. Planck's
noise/beam information is obtained from {\sf
http://www.rssd.esa.int/Planck}). The fraction of sky $f_{\rm sky}$ is
set to be 1 in this analysis.

\begin{figure}
\begin{center}
\includegraphics[width=8.7cm]{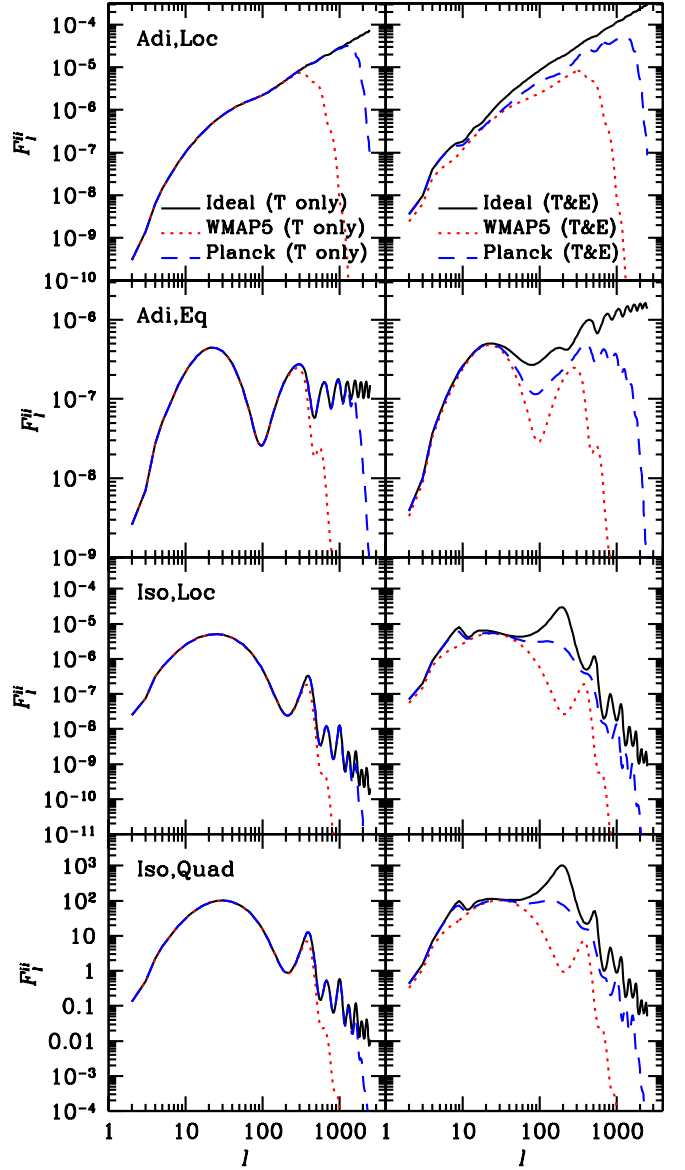}
\caption{Diagonal components of the Fisher matrix $F_l^{ii}$
(eq.[\ref{eq:fisher}]). From top to bottom, the local-type adiabatic
(Adi,Loc), the equilateral-type adiabatic (Adi,Eq), the local-type
isocurvature (Iso,Loc), and the quadratic-type isocurvature (Iso,Quad)
components are plotted. Left panels are for T map only, but right
panels are for T\&E maps.  Noise/Beam is for Ideal (solid), WMAP5
(dotted) and Planck observations (dashed).}
\label{fig:fisauto}
\end{center}
\end{figure}

Figure \ref{fig:fisauto} shows the diagonal component of the Fisher
matrix $F_l^{ii}$ (eq.[\ref{eq:fisher}]).  It represents the square of
signal-to-noise ratio $(S/N)^2$ for $i$-th NG component at $l$ without
correlations among different NG components.  The adiabatic components
have an increasing trend of $S/N$ at higher $l$. The majority of the
signal of the isocurvature components in temperature maps come from
the large-angular scale ($l<100$), where isocurvature perturbations
produce larger CMB fluctuations than adiabatic perturbations. A phase
difference in acoustic oscillations between adiabatic and isocurvature
modes provides a distinct signature seen around $l\sim 300$, which is
important particularly when polarization maps are included. Table
\ref{tab:sprim} lists the values of the diagonal components of the
Fisher matrix summed over $l$ up to 2500, at
which Planck estimates are enough saturated.

\begin{table}
\caption{Diagonal components of the Fisher matrix $F^{ii}$ summed up
to $l=2500$.  The different noise/beam for WMAP5, Planck, and Ideal
are considered.}
\begin{center}
\begin{tabular}{ccccc}
  \hline\hline
       & Adi,Loc & Adi,Eq & Iso,Loc & Iso,Quad \\ 
\hline
WMAP5 (T only)  & 2.7$\times 10^{-3}$ & 7.2$\times 10^{-5}$ 
    & 2.9$\times 10^{-4}$ & 6.8$\times 10^3$ \\
Planck (T only)  & 3.7$\times 10^{-2}$ & 2.3$\times 10^{-4}$ 
    & 3.1$\times 10^{-4}$ & 7.6$\times 10^3$ \\
Ideal (T only) & 9.0$\times 10^{-2}$ & 3.4$\times 10^{-4}$ 
    & 3.1$\times 10^{-4}$ & 7.7$\times 10^3$ \\
\hline
WMAP5 (T\&E)  & 3.0$\times 10^{-3}$ & 7.5$\times 10^{-5}$ 
    & 3.1$\times 10^{-4}$ & 7.2$\times 10^3$ \\
Planck (T\&E)  & 5.8$\times 10^{-2}$ & 4.8$\times 10^{-4}$ 
    & 8.9$\times 10^{-4}$ & 2.6$\times 10^4$ \\
Ideal (T\&E) & 3.6$\times 10^{-1}$ & 2.9$\times 10^{-3}$ 
    & 3.9$\times 10^{-3}$ & 1.3$\times 10^5$ \\
\hline
\end{tabular}
\end{center}
\label{tab:sprim}
\end{table}

\begin{figure*}
\includegraphics[width=9.17cm]{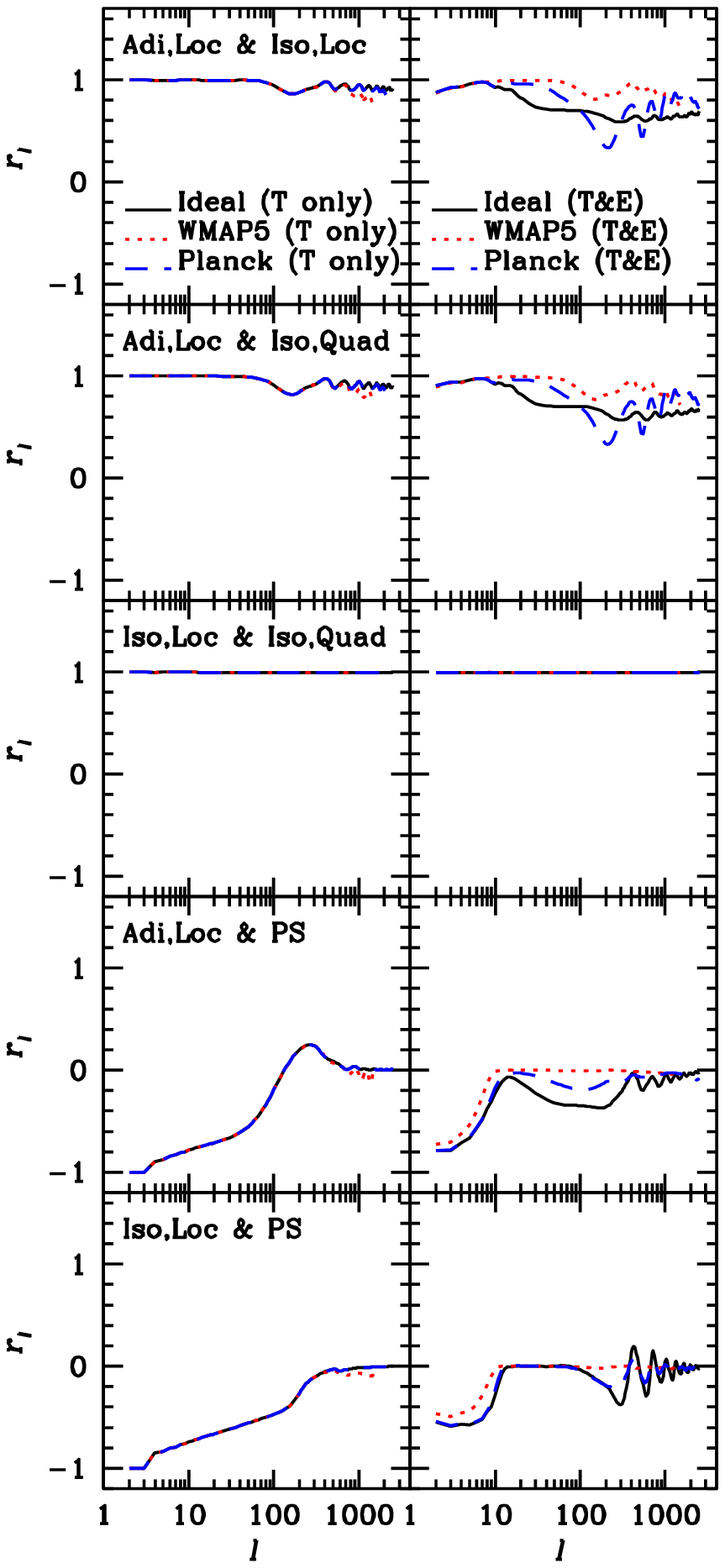}
\includegraphics[width=8cm]{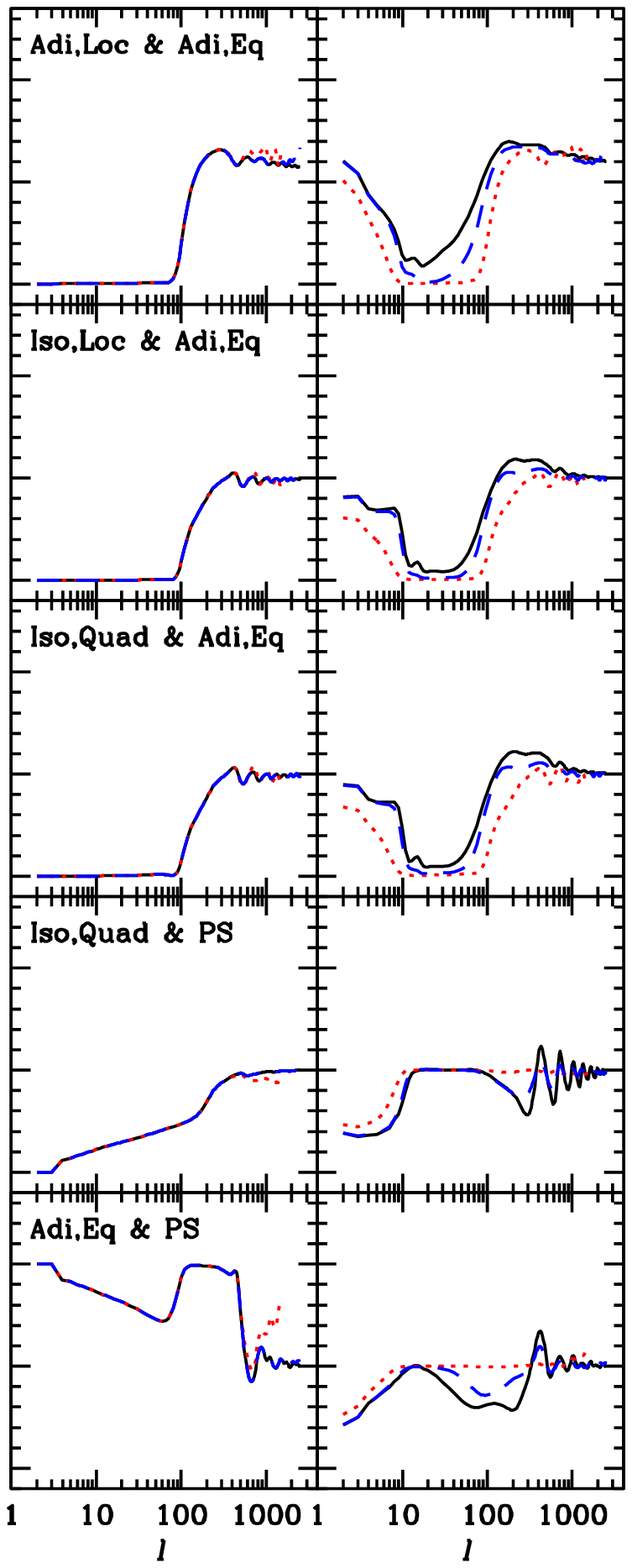}
\begin{center}
\caption{Cross-correlation coefficients $r_l^{ij}\equiv
F_l^{ij}/(F_l^{ii}F_l^{jj})^{1/2}$ where $i$ and $j$ denote the
local-type adiabatic (Adi,Loc), the local-type isocurvature (Iso,Loc),
the equilateral-type adiabatic (Adi,Eq), the quadratic-type
isocurvature (Iso,Quad), and the point source (PS) components.  Left
of each panel is temperature T only; right is T+E, including E
polarization. Noise/Beam is for Ideal (solid), WMAP5 (dotted) and
Planck observations (dashed).}
\label{fig:slcorr}
\end{center}
\end{figure*}

Figure \ref{fig:slcorr} shows the cross-correlation coefficient $r_l$
defined as $r_l\equiv F_l^{ij}/(F_l^{ii}F_l^{jj})^{1/2}$.  The
local-type adiabatic and isocurvature components are strongly
correlated, but the phase difference of acoustic oscillations
weakens the correlation, as seen especially around $l\sim 200$.  The
correlation between the local-type and the equilateral-type components
becomes weak at $l>100$. The two isocurvature components are almost
completely correlated over all scales.  The correlation with the point
source component is very weak for $l>100$.

\begin{figure*}
\includegraphics[width=14cm]{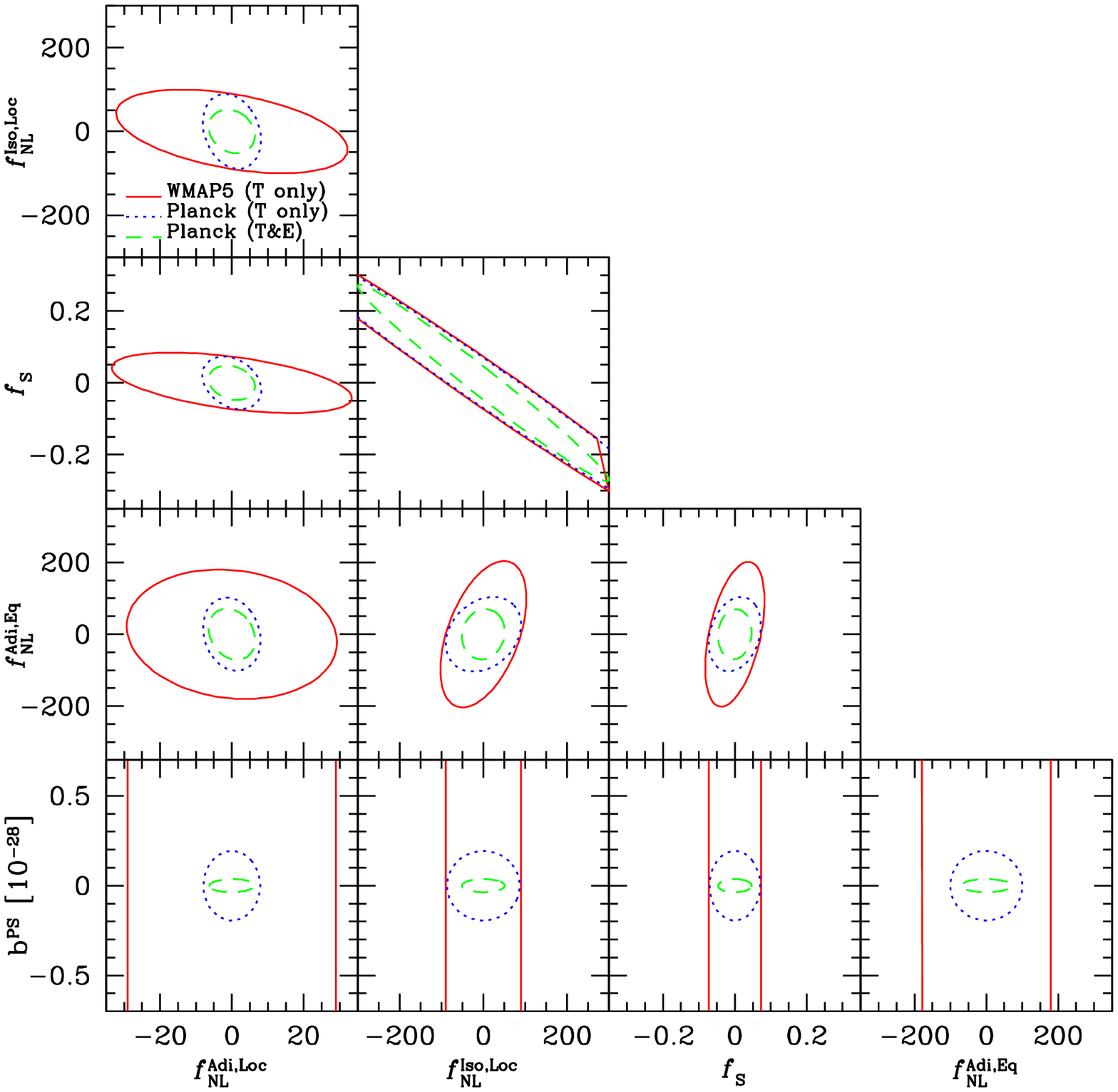}
\begin{center}
\caption{1$\sigma$ error contours of a pair of NG parameters 
expected from WMAP5 T only (solid circles), Planck T only (dotted circles), 
and Planck T\&E (dashed circles). The rest of NG parameters other
than two plotted are fixed to be zero.}
\label{fig:jointerr}
\end{center}
\end{figure*}

Figure \ref{fig:jointerr} shows 1$\sigma$ error contours
(Cram\'{e}r-Rao bound) for a pair of NG parameters for WMAP5 (T only),
Planck (T only), and Planck (T\&E). The errors expected from WMAP5
(T\&E) is almost same as those from WMAP5 (T only). The rest of NG
parameters other than two plotted are fixed to be zero. The local-type
adiabatic and isocurvature components are correlated with the
correlation coefficient $r=0.43$ for WMAP5, $r=0.23$ for Planck T
only, $r=0.20$ for Planck T\&E when $l$ is summed up to 2500. We see
that the local-type and the quadratic-type scale-invariant
isocurvature components are difficult to be differentiated even using
Planck data. The local-type and the equilateral-type adiabatic
components are weakly correlated ($r=0.12$ for WMAP5, $r=0.17$ for
Planck T only, $r=0.22$ for Planck T\&E), which is consistent with the
previous work \citep{BCZ04}. The point source component is almost
uncorrelated with the other primordial components ($r<0.08$ for WMAP5
and $r<0.03$ for Planck), which is consistent with the previous work
\citep{KomSpe01}.  Table \ref{tab:limit} lists the errors of the NG
parameters without and with correlations among all of other parameters
except for the quadratic-type isocurvature component.  Polarization
maps are found to be very important to constrain the isocurvature NG
as well as adiabatic NG.  The increase of the errors due to the
correlations mainly between adiabatic and isocurvature modes is
20-30\% for WMAP5, but less than 5\% for Planck observations.

\begin{table}
\caption{1$\sigma$ errors of each NG parameter from the joint analysis
of all of the NG components except for the quadratic-type isocurvature
components.  The values without parentheses denote the limits with
other parameters fixed, but those with parentheses denote the limits
including the correlations such that the other parameters are
marginalized.}
\begin{center}
\begin{tabular}{ccccc}
  \hline\hline
   & WMAP5 & & \multicolumn{2}{c}{Planck} \\ 
   \cline{2-2}
   \cline{4-5}
   & \raisebox{-1ex}{T only} &
   & \raisebox{-1ex}{T only} & \raisebox{-1ex}{T\&E} \\
\hline
$\Delta f_{\rm NL}^{\rm Adi,Loc}$ &  19 (23) & & 5.2 (5.5) & 4.1 (4.3) \\
$\Delta f_{\rm NL}^{\rm Iso,Loc}$ &  59 (82) & & 57 (62) & 34 (35)  \\
$\Delta f_{\rm NL}^{\rm Adi,Eq}$  &  117 (149) & & 66 (71) & 46 (48)  \\
$\Delta b^{\rm PS} [10^{-28}]$  & 438 (441) & & 0.13 (0.13) & 0.024 (0.024) \\
  \hline
\end{tabular}
\end{center}
\label{tab:limit}
\end{table}

\section{Summary}
\label{sec:summary}

We have presented a detailed analysis of the possibility of extracting
information about non--Gaussianity from various inflationary
models. We consider four different-type primordial NG models:
local-type adiabatic, equilateral-type adiabatic, local-type
isocurvature, and quadratic-type isocurvature models together with
point source contamination. The adiabatic and the isocurvature modes
are correlated, but the difference in the phase of the corresponding
acoustic oscillations breaks the degeneracy.  The local-type and
quadratic-type scale-invariant isocurvature components are difficult
to separate even using Planck data.  The correlation between the
local-type and the equilateral-type adiabatic modes is weak.  The
point source (white noise) contamination does not pose a threat as it
is uncorrelated with any of the $f_{\rm NL}$ parameters, although a
high-resolution experiment will be more suited to get rid of such
contamination.  Our results are based on noise models from WMAP and
Planck and we compare them to ideal noise-free and all-sky reference
observations.  The increase of the error for the non-Gaussian
parameters due to the correlations is 20-30\% for WMAP5 and 5\% for
Planck.

Secondary anisotropies other than point sources can
contaminate the estimation of primordial non--Gaussianity. The
cross-contamination of various inflationary contributions against
secondaries such as Sunyaev-Zeldovich effect (SZ) or Integrated
Sachs-Wolfe effect (ISW) which are potentially observable with Planck
data will be present elsewhere.

\section*{Acknowledgments}
CH acknowledges support from a JSPS (Japan Society for the
Promotion of Science) fellowship. DM acknowledges financial support
from an STFC rolling grant at the University of Edinburgh.

{}

\end{document}